\documentclass[10pt,dvipsnames,5p,british]{elsarticle}
\biboptions{numbers,sort&compress}
\usepackage[british]{babel}
\usepackage{amsmath,amsfonts,amssymb}
\usepackage{color}
\usepackage[dvipsnames]{xcolor}
\usepackage{braket}
\usepackage{graphicx,tikz,pgfplots}
\usepackage[final]{hyperref} 
\hypersetup{
	colorlinks=true,       
	linkcolor=red,        
	citecolor=red,        
	filecolor=magenta,     
	urlcolor=blue         
}
\usepackage{lipsum}
\usepackage{multicol}
\usepackage{float}
\usepackage{caption}
\usepackage{subcaption}
\numberwithin{equation}{section}
\numberwithin{figure}{section}
\numberwithin{table}{section}
\pgfplotsset{compat=1.17}
\usetikzlibrary{spy}
\usepgfplotslibrary{groupplots}
\usetikzlibrary{external}
\usepgfplotslibrary{external}

\begin{document}
\title{Confinement in the tricritical Ising model}
\author[1]{M. Lencs{\'e}s}
\author[2]{G. Mussardo}
\author[1,3]{G. Tak{\'a}cs}
\address[1]{BME-MTA Statistical Field Theory ’Lend{\"u}let’ Research Group,\\ Department of Theoretical Physics, Budapest University of Technology and Economics,\\ 1111 Budapest, M{\H u}egyetem rkp. 3, Hungary}
\address[2]{SISSA and INFN, Sezione di Trieste,\\ via Bonomea 265, I-34136, Trieste, Italy}
\address[3]{MTA-BME Quantum Dynamics and Correlations Group, E{\"o}tv{\"o}s Lor{\'a}nd Research Network\\ 1111 Budapest, M{\H u}egyetem rkp. 3, Hungary}

\date{February 4, 2022}

\begin{abstract}
 
  We study the leading and sub-leading magnetic perturbations of the thermal $E_7$ integrable deformation of the tricritical Ising model. In the low-temperature phase, these magnetic perturbations lead to the confinement of the kinks of the model. The resulting meson spectrum can be obtained using the semi-classical quantisation, here extended to include also mesonic excitations composed of two different kinks.  An interesting feature of the integrable sub-leading magnetic perturbation of the thermal $E_7$ deformation of the model is the possibility to swap the role of the two operators, i.e. the possibility to consider the model as a thermal perturbation of the integrable $\mathcal{A}_3$ model associated to the sub-leading magnetic deformation. 
  Due to the occurrence of vacuum degeneracy unrelated to spontaneous symmetry breaking in $\mathcal{A}_3$, the confinement pattern shows novel features compared to previously studied models. Interestingly enough, the validity of the semi-classical description in terms of the $\mathcal{A}_3$ endpoint extends well beyond small fields, and therefore the full parameter space of the joint thermal and sub-leading magnetic deformation is well described by a combination of semi-classical approaches. All predictions are verified by comparison to finite volume spectrum resulting from truncated conformal space. 
  
\end{abstract}
\maketitle

\section{Introduction}
Confinement plays a fundamental role in strong interactions as the mechanism ensuring that colour charged particles (quarks and gluons) are never directly observed in Nature: the physical particle spectrum consists solely of colour singlet mesons and baryons \cite{1974PhRvD..10.2445W}. For mesons, confinement can be described in terms of a quark-antiquark potential that rises linearly for long distances, leading to a force independent of separation. Besides its importance in high-energy physics, confinement also arises in low dimensional quantum field theories describing the continuum limit of quantum spin chains as originally found by McCoy and Wu in the case of the Ising model \cite{1978PhRvD..18.1259M}. As shown in \cite{1996NuPhB.473..469D, Delfino:1997ya,2011JSMTE..01..002M}, in this case the origin of this confinement can be traced back to the non-locality of the elementary excitations of the system (given, in the ferromagnetic phase, by domain walls a.k.a. "kinks" with different topological charges interpolating between regions corresponding to different vacua) with respect to the operator which perturbs the system.  

Recent interest in confinement was spurred by the discovery that it leads to drastic change of non-equilibrium dynamics in quantum spin chains, first demonstrated for the Ising model in  \cite{2017NatPh..13..246K}. The effect, known as dynamical confinement, has also been experimentally observed in quantum simulators \cite{2019arXiv191211117T} and demonstrated using quantum computers \cite{2021NatSR..1111577V}.

Here we investigate confinement in the Tricritical Ising Model (TIM), which has several novel interesting features compared to the previously considered Ising and Potts cases. Namely, the phase pattern of the model is much more involved, allowing a smooth interpolation between limits with very different vacuum structure, and correspondingly, very different kink structure. A related aspect is the presence of {\em two} independent magnetisation operators, both of which can induce confinement of the domain walls in the ferromagnetic phase. In a recent paper \cite{2021arXiv210909767C} we showed that this also results in interesting new aspects of Kramers--Wannier duality and the form factor bootstrap. Recent studies of kink scattering also indicate significant difference between the confinement dynamics of the Ising/Potts and the tricritical Ising model \cite{2020arXiv201207243M}.

A novel feature in the TIM emerges if one considers a combination of thermal and sub-leading magnetic deformations, associated respectively to the $E_7$ and $\mathcal{A}_3$ scattering theories: such combined perturbation of the TIM interpolates between endpoints with two degenerate vacua but of different nature (cf. Fig. \ref{fig:GL}). The $E_7$ endpoint of purely thermal perturbation has two symmetric vacua  related to the spontaneous breaking of $\mathbb{Z}_2$ symmetry while the $\mathcal{A}_3$ endpoint associated to purely subleading magnetic perturbation presents two degenerate but asymmetric vacua with a peculiar kink structure since the two vacua are not related to any spontaneous symmetry breaking. Hence, the perturbation of the tricritical Ising fixed point by the combination of thermal and sub-leading magnetisation operators can be described in terms of two qualitatively different confinement dynamics.

What are the consequences of the confinement of the kinks? The most important one is the appearance of a new spectrum of mesonic excitations. 
Indeed, when the degeneracy between the vacua is lifted by the presence of an external magnetic field, this leads to an energy contribution proportional to the separation of domain walls, giving rise to the confinement of the domain walls in those channels for which the induced force is attractive \cite{1996NuPhB.473..469D}. Hence, in these statistical field theories the role of the quarks is played by the kinks, while the colour charge of QCD corresponds to the topological charge carried by them. A classification of kink systems and their topological charges for simple perturbed conformal field theories, together with a description of their kink systems and patterns of confinement can be found in \cite{2009JPhA...42D4022M}. 

In a semi-classical approximation, the meson spectrum is simply described by bound states of a two-kink ("quark-antiquark") system in the linear potential induced by the magnetic field \cite{2005PhRvL..95y0601R}. For the Ising model where the broken symmetry classifying the topological charges is $\mathbb{Z}_2$, the nature of the meson spectrum is well understood beyond the semi-classical regime by using the Bethe-Salpeter equation \cite{Ising_analytic,2005PhRvL..95y0601R,2006hep.th...12304F,2017NuPhB.923..508R,2009JPhA...42D4025R}. For the 3-state Potts model where the symmetry is instead the permutation group $S_3$, switching on a magnetic field in the ferromagnetic phase leads to a confined spectrum containing also "baryonic" bound triplets of the "quarks" besides the "quark-antiquark" mesons \cite{2008NuPhB.791..265D,2009JSMTE..11..007L}. The meson spectrum in the Potts model can be extracted by means of an improved version of the semi-classical approximation \cite{2010JPhA...43w5004R}, while the spectrum of baryons can be described using a quantum mechanical three-body problem with a linear interaction potential \cite{2015JSMTE..01..010R}. We note that the semi-classical approximation to the meson spectrum is also very efficient for one-dimensional lattice magnets a.k.a. spin chains and ladders \cite{2010JSMTE..07..015R,2017PhRvB..96e4423B,2018EL....12137001R,2020JSMTE2020i3106L,2020PhRvB.102a4426R,2021arXiv211009472L}. 

In the case of the TIM perturbed by the combined thermal and sub-leading magnetisation operators,  the different vacuum structures of the two endpoints of this theory result in two independent and distinct semi-classical descriptions of the same meson spectrum. While usually the semi-classical approximation only works when the corresponding confining field is weak, as a matter of fact for the TIM one of the semi-classical approaches is quite accurate even in the strong field regime, a circumstance which permits an efficient control of the spectrum in a broad range of values of the coupling constants. To validate the theoretical predictions for the confinement dynamics and to test the accuracy of the semi-classical predictions, we use a truncated Hamiltonian method which permits an efficient numerical control of the spectrum (this method was previously applied in the context of the Ising \cite{2006hep.th...12304F} and Potts models \cite{2015JHEP...09..146L}).

The outline of the work is as follows. In Section \ref{sec:confinement} we recall and extend the semi-classical description of the meson spectrum resulting from confinement. Section \ref{sec:confinement_tricritical} describes the application of the semi-classical method to the tricritical Ising model, and its predictions are validated by comparing to the truncated conformal space approach in Section \ref{sec:tcsa_comparison}. Finally we sum up our conclusions in Section \ref{sec:conclusions}.

\section{Confinement and semi-classical meson spectrum}\label{sec:confinement}

Here we summarise the semi-classical description of confinement, based on the approach developed by Rutkevich \cite{2005PhRvL..95y0601R}, and we also generalise it for particles composed of two kinks of different species. The semi-classical motion of the two kinks is described by the following Hamiltonian
\begin{equation}
    H=\omega_a(p_1) + \omega_b(p_2) + \Delta \mathcal{E} |x_1 - x_2|
\end{equation}
where $\Delta \mathcal{E}$ is the "string tension" responsible for confinement, and the dispersion relations are 
\begin{equation}
    \omega_a(p) = \sqrt{m_a^2 + p^2}\,,
\end{equation} 
which admits the familiar rapidity parameterisation $p=m_a\sinh\theta$ and $\omega_a(p)=m_a\cosh\theta$. Note that in this approach, the meson wave functions are approximated with a linear combination of the unperturbed ($\Delta\mathcal{E}=0$) two-kink scattering states, a crucial assumption of the semi-classical description.

Assuming first that the two kinks are of the same species $a=b$ (in particular, $m_a=m_b=m$ and therefore $p_1=-p_2=p$), for well-separated kinks with $x\gg m^{-1}$ (right transport region) the wave function in the centre-of-mass frame satisfies the following Schrödinger equation:
\begin{equation}
    \Big[2\omega(\hat{p})-E_n+\Delta \mathcal{E}|x|\Big]\phi^+_n(x) = 0\quad,\quad\hat{p}=-i\partial_x\ ,
\end{equation}
where $E_n$ is the energy of the bound state. In momentum space it takes the form
\begin{equation}
    \Big[2\omega(p)-E_n+i\Delta \mathcal{E}\, \partial_p\Big]\phi_n(p) = 0
\end{equation}
with the solution
\begin{equation}
\label{eq:idWF}
    \phi^+_n(x) = \int_{-\infty}^{\infty} \frac{dp}{2\pi}\exp\left\{\frac{i\left[f(p)-p \, E_n\right]}{\Delta \mathcal{E}} +ipx \right\}
\end{equation}
where 
\begin{equation}
    f(p)=\int_{0}^{p}2\omega(q)dp=m^2\left(\theta+\frac{\sinh2\theta}{2}\right)
\end{equation}
having used the rapidity parameter $\theta$ defined by $p=m\sinh\theta$. For sufficiently large values of $E_n$ the integral \eqref{eq:idWF} can be evaluated using a saddle point approximation
\begin{equation}
    \phi^+_n(x) \approx \mathcal{C}(\theta_n) e^{i m x \sinh\theta_n} + \mathcal{C}^{*}(\theta_n) e^{-i m x \sinh\theta_n}\,,
\end{equation}
with contributions from two saddle points located in the rapidity variable at $\pm \theta_n$ which is related to the energy of the state by $E_n=2m\cosh\theta_n$. The amplitudes are given by
\begin{equation}
    \mathcal{C}(\theta_n) = \exp 
    \left[ \frac{i m^2}{\Delta\mathcal{E}}
    \left(\theta_n-\displaystyle\frac{\sinh2\theta_n}{2}\right)
    +\frac{i \pi}{4} \right]
\end{equation}
and they are related by the two-kink scattering matrix $S(\theta_1-\theta_2)$ in the following way:
\begin{equation}
    \mathcal{C}(\theta_n)=S(2\theta_n)\mathcal{C}^*(\theta_n)\,,
\end{equation}
which leads to the quantisation condition
\begin{equation}
\label{eq:idKquant}
    \sinh 2\theta_n - 2\theta_n = \frac{\Delta \mathcal{E}}{m^2}\left[ 2\pi\left(n+\frac{1}{4}\right) + i \log S(2\theta_n)\right]
\end{equation}
The above derivation must be changed slightly when the two kinks are not identical. Now the Schrödinger equation for the relative motion has the form
\begin{equation}
\Big(\omega_a(p)+\omega_b(p)-E_n+\Delta \mathcal{E}x \Big)\phi_n^+(x)=0
\end{equation}
The saddle-point evaluation of the wave function leads to 
\begin{equation}
    \phi^+_n(x) \approx \mathcal{C}(\theta_n) e^{i m_a x \sinh\theta_n} + \mathcal{C}^{*}(\theta_n) e^{-i m_a x \sinh\theta_n}
\end{equation}
where
\begin{align}
    \mathcal{C}(\theta_n) = \exp \Bigg[ 
    &i\frac{m_a^2}{2\Delta \mathcal{E}}\left(\theta_n-\frac{\sinh2\theta_n}{2}\right)
    \nonumber\\
    -&i\frac{m_b^2}{2\Delta \mathcal{E}}\left(\tilde{\theta}_n-\frac{\sinh2\tilde{\theta}_n}{2}\right) +i\frac{ \pi}{4} \Bigg]
\end{align}
with the momenta of the two kinks written with rapidities defined as  $p=m_a\sinh\theta_n=-m_b\sinh\tilde{\theta}_n$. Matching the two plane wave components results in the condition
\begin{equation}
    \mathcal{C}(\theta_n)=\pm S_{ab}(\theta_n-\tilde{\theta}_n)\mathcal{C}^*(\theta_n)\ ,
\end{equation}
where the sign choice corresponds to a wave-function symmetric/antisymmetric under the exchange of the constituent kinks (note that in the case of identical particles the wave-function is automatically symmetric). This leads to the quantisation condition
\begin{align}
& m_a^2\left(\frac{1}{2}\sinh2\theta_n-\theta_n\right)- 
m_b^2\left(\frac{1}{2}\sinh2\tilde{\theta}_n-\tilde{\theta}_n\right)
\nonumber\\
&=\Delta\mathcal{E}\left[2\pi\left( \frac{n}{2} +\frac{1}{4}\right)+i\log S_{ab}(\theta_n-\tilde{\theta}_n)\right]
\nonumber\\
&\textrm{where }\tilde{\theta}_n=-\sinh^{-1}\left(\frac{m_a\sinh\theta_n}{m_b}\right)
\qquad
\label{eq:nonidKquant}
\end{align}
The mass of the bound state is then given by $M^{a,b}_n=m_a \cosh\theta_n+m_b\cosh\tilde{\theta}_m$. For the case $m_a=m_b=m$ the rapidities satisfy $\tilde{\theta}_n=-\theta_n$ and replacing the quantum number $n$ by $2n$ (corresponding to a symmetric wave-function), the above equation reduces to \eqref{eq:idKquant}. Note that the equations \eqref{eq:idKquant} and \eqref{eq:nonidKquant} depend on the dimensionless ratio $\zeta=\Delta\mathcal{E}/m^2$, where $m$ is the lower mass entering \eqref{eq:nonidKquant}. As well known, the semi-classical approximation is limited to suitably small values of $\zeta$ as well as to large enough values of excitation number $n$ \cite{2005PhRvL..95y0601R}.

\section{Semi-classical meson spectra in the tricritical Ising model}\label{sec:confinement_tricritical}
\begin{figure*}[ht!]
    \centering
    \includegraphics[width=\textwidth]{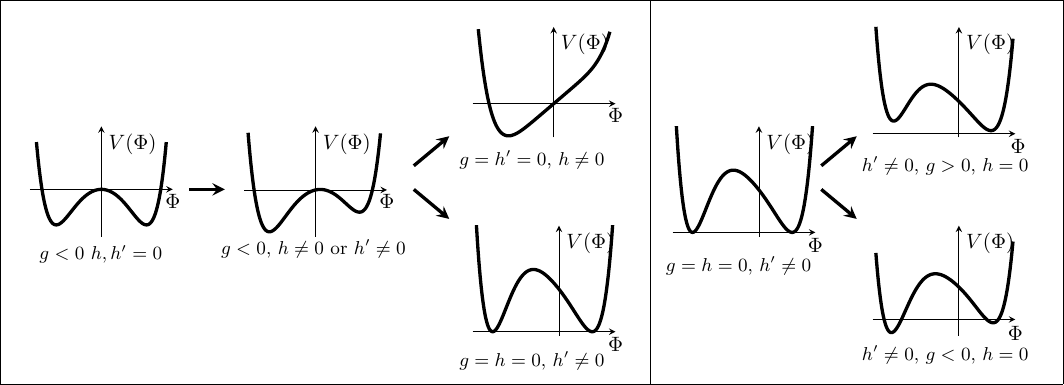}
    \caption{Qualitative shapes of the  Ginzburg--Landau potential $V(\Phi)$ for the perturbations considered in this paper.}
    \label{fig:GL}
\end{figure*}

We apply the above semi-classical approach to perturbations of the tricritical Ising model, which is a minimal model of conformal field theory with central charge $c=7/10$ \cite{BPZ1984}. The perturbing operators we consider hereafter are the thermal perturbation $\epsilon$ with conformal weights $h_\epsilon=\bar{h}_\epsilon=1/10$, the leading magnetisation operator $\sigma$ with $h_\sigma=\bar{h}_\sigma=3/80$, and the subleading magnetisation operator $\sigma'$ with $h_{\sigma'}=\bar{h}_{\sigma'}=7/16$. The most general action that we will consider in the following can be written as 
\begin{align}
    \mathcal{A}\left[g,h,h'\right]\,\, =\, &\mathcal{A}_\mathrm{CFT}  \nonumber\\
    & +\int d^2x \left( g\:\epsilon(x)+ h\:\sigma(x) +h'\:\sigma'(x) \right)
\end{align}
where $\mathcal{A}_\mathrm{CFT}$ denotes the action for the fixed point conformal field theory. As well known \cite{Zamolodchikov:1986db,Mussardo:2020rxh}, the class of universality of tricritical Ising can put in correspondence with a Landau--Ginzburg model based on a scalar field $\Phi(x)$ and a potential $V(\Phi)$ of sixth degree. For the cases considered in this paper, the qualitative shapes of the potential $V(\Phi)$ are shown in Fig.~\ref{fig:GL}.

\subsection{Confinement in the $E_7$ model induced by magnetisation perturbations}\label{subsec:e7_confinement}

\begin{figure}
    \centering
    \includegraphics[]{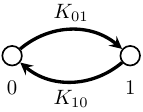}
    \caption{Kink structure for the $E_7$ model.}
    \label{fig:E7kinks}
\end{figure}

The pure thermal perturbation $\mathcal{A}\left[g,0,0\right]$ gives rise to a theory with a $\mathbb{Z}_2$ invariance, which is spontaneously broken in the ferromagnetic phase $g<0$, while $g>0$ corresponds to the paramagnetic phase. This perturbation is integrable and its spectrum is described by the $E_7$ scattering theory \cite{ChristeMussardo1990,FateevZamolodchikov1990}. In the low-temperature phase, characterised by two equivalent vacuum states, the exact spectrum contains three kink excitations $A_1$, $A_3$, and $A_6$ corresponding to domain walls between the two vacua and four breather states $A_2$, $A_4$, $A_5$, and $A_7$ which correspond to kink-antikink bound states (a.k.a. breathers). The kink excitations interpolate between the two vacua as shown in Fig. \ref{fig:E7kinks} where the generic kink doublet $K_{01}$, $K_{10}$ can be any of $A_1$, $A_3$, and $A_6$, and the latter two can be considered as excited versions of the fundamental kink $A_1$. We denote the masses of the excitations $A_k$ with $m_k$; their ratios are exactly known  in terms of the pole structure of the $E_7$ scattering amplitudes \cite{ChristeMussardo1990,FateevZamolodchikov1990}, and the mass gap can also be exactly computed in terms of the coupling constant \cite{Fateev1994}:
\begin{equation}
  m_1=3.745372836\dots\cdot|g|^{5/9}  \,.
  \label{eq:e7gap}
\end{equation}
To induce confinement of the kink excitations, we consider the ferromagnetic regime and add the magnetic fields which break $\mathbb{Z}_2$ explicitly and result in the full action $\mathcal{A}\left[g,h,h'\right]$ (with $g<0$). With these perturbations switched on the double degeneracy of the vacua is lifted and, 
to first order in $h$ and $h'$, the relative energy density of the false vacuum is
\begin{equation}
	\label{eq:relen}
	\Delta \mathcal{E} = \left|2 h \braket{\sigma}+2 h' \braket{\sigma'}\right|
\end{equation}
where 
\begin{align}
\braket{\sigma}&=1.5927\dots \cdot(-g)^{1/24} \nonumber\\
\braket{\sigma'}&=2.45205\dots \cdot(-g)^{35/72}    
\end{align} 
are the symmetry-breaking expectation values of the leading and subleading magnetisation operators in the $E_7$ model $\mathcal{A}\left[g,0,0\right]$~\cite{FLZZ}. The meson spectrum coming from the original two-kink states $|A_1A_1\rangle$ is determined by Eqn. \eqref{eq:idKquant} with $m$ identified with the mass $m_1$ of $A_1$, and the exact two-kink scattering amplitude given by
\begin{align}
S_{11}(\theta)=&-f_{2}(\theta)f_{10}(\theta)
\\
&f_\alpha(\theta)\,\equiv \,\frac{\tanh\frac{1}{2}\left(\theta + i\pi\frac{\alpha}{18}\right)}{\tanh\frac{1}{2}\left(\theta - i \pi \frac{\alpha}{18}\right)}\,.\nonumber
\end{align}
The meson masses can be computed from the solutions of  \eqref{eq:idKquant} as $M_n=2 m_1\cosh\theta_n$. Note that the meson spectrum only depends on the dimensionless parameter $\zeta_{1}=\Delta\mathcal{E}/m_1^2$ and,  in particular, the semi-classical prediction of this quantity does not depend separately on the two magnetic fields $h$ and $h'$.  

Concerning the stability of these mesons, those with masses above the two-particle threshold are expected to be unstable and to decay. In our case the two-particle threshold is given by $2m_2(\zeta_1)$, where $$m_2(\zeta_1)=2 m_1 \cos(5\pi/18)+O(\zeta_1^2)$$ is the mass of the original particle $A_2$ as a function of the perturbation parameter $\zeta_1$. It is interesting to notice that for the leading magnetic perturbation all the mesons become unstable around $\zeta_1\approx 0.1$, while for the subleading magnetic perturbation the four mesons shown in Fig.~\ref{fig:mesonsigma} are all stable.

\subsection{Confinement in the RSOS $\mathcal{A}_3$ model induced by the thermal perturbation}\label{subsec:a3_confinement}

\begin{figure}
    \centering
    \includegraphics[]{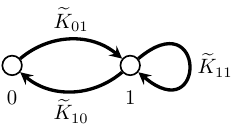}
    \caption{Kink structure for the $\mathcal{A}_3$ model.}
    \label{fig:A3kinks}
\end{figure}

It is worth to underline that the perturbation of the tricritical Ising model by its subleading magnetisation operator
\begin{align}
    \mathcal{A}\left[0,0,h'\right]\,=\, &\mathcal{A}_\mathrm{CFT} + h'\int d^2x\,  \sigma'(x)
    \label{suuub}
\end{align}
is integrable and describes the scaling limit of the dilute RSOS $\mathcal{A}_3$ lattice model \cite{2002JPhA...35.1597S}. The dynamics is independent of the sign of $h'$, which we choose positive for definiteness. In this case the quantum field theory (\ref{suuub})  has two {\em inequivalent} ground states $\ket{0}$ and $\ket{1}$: the fundamental excitations of the model are the kinks $|\widetilde{K}_{01}(\theta)\rangle$, $|\widetilde{K}_{10}(\theta)\rangle$ and $|\widetilde{K}_{11}(\theta)\rangle$, where the underline indices correspond to the vacua interpolated by the kinks as shown in Fig. \ref{fig:A3kinks}. Notice the absence from the spectrum of the state $|\widetilde{K}_{00}(\theta)\rangle$, a fact which further confirms the inequivalence of the two vacua. 

The two-kink scattering amplitudes $S_{ac}^{bd}(\theta_1-\theta_2)$ describing the processes 
\begin{equation}
    \widetilde{K}_{ab}(\theta_1)+\widetilde{K}_{bc}(\theta_2)\rightarrow \widetilde{K}_{ad}(\theta_1)+\widetilde{K}_{dc}(\theta_2)
\end{equation}
are known exactly \cite{1992PhLB..274..367C,1992IJMPA...7.5281C}. The mass of the kink excitations is related to the coupling constant by 
\begin{equation}
    m_K=4.927791224\dots\cdot|h'|^{8/9}\,.
\label{eq:a3gap}
\end{equation}
For this theory the kink excitations can be confined by switching on the coupling $g$ of the thermal perturbation $\epsilon$. Interestingly enough, for $g<0$ this leads to an independent semi-classical description for the spectrum of the model $\mathcal{A}\left[g,0,h'\right]$. In such a case, it is the ground state $\ket{1}$ which becomes the false vacuum, its energy density relative to the true vacuum $\ket{0}$ given by
\begin{equation}
\label{eq:endensA3}
\widetilde{\Delta \mathcal{E}} = |g(\braket{\varepsilon}_1-\braket{\varepsilon}_0)|,
\end{equation}
where~\cite{FLZZ}
\begin{align}
\braket{\varepsilon}_1&=2.0445\dots \left|h'\right|^{8/45}\,,\nonumber\\
\braket{\varepsilon}_0&=-0.78093\dots \left|h'\right|^{8/45}\,.
\end{align}
Therefore the meson spectrum (coming from the confinement of two-kink states  $|\widetilde{K}_{01}\widetilde{K}_{10}\rangle$) can be obtained from \eqref{eq:idKquant} substituting $m_K$ for $m$ and the kink scattering amplitude
\begin{align}
    S_{00}^{11}(\theta)=&
    \prod_{\alpha\in\Pi_1} \frac{\displaystyle\sinh\left(\frac{9}{10}\theta+i\pi\alpha \right)}{\displaystyle\sinh\left(\frac{9}{10}\theta-i\pi\alpha \right)}
    \prod_{\alpha\in\Pi_2} \frac{\displaystyle\sinh\frac{1}{2}\left(\vphantom{\frac{1}{2}}\theta+i\pi\alpha \right)}{\displaystyle\sinh\frac{1}{2}\left(\vphantom{\frac{1}{2}}\theta-i\pi\alpha \right)}
    \nonumber \\
    \Pi_1&=\left\{-\frac{1}{5},\frac{3}{10}\right\}
    \quad \Pi_2=\left\{\frac{2}{9},-\frac{8}{9},\frac{7}{9},-\frac{1}{9}\right\}
\end{align}
for $S(\theta)$. Let's notice that the kink $\widetilde{K}_{11}$ disappears from the spectrum of stable particle states, and so the stability condition of the mesons is given in this case by $\widetilde{M}_n<2\widetilde{M}_1$, with the threshold determined by the mass $\widetilde{M}_1$ of the lightest meson. Since $\widetilde{M}_1>2m_K$, mesons with masses below $4m_K$ are always stable. 

When $g>0$, it is the ground state $\ket{0}$ which becomes instead the false vacuum and therefore one naively expects the confinement of two-kink states  $|\widetilde{K}_{10}\widetilde{K}_{01}\rangle$. However, the interaction of the kinks allows for the process $|\widetilde{K}_{10}\widetilde{K}_{01}\rangle  \rightarrow |\widetilde{K}_{11}\widetilde{K}_{11}\rangle$, and this transition is energetically favoured compared to the preservation of the false vacuum between the two kinks. Consequently, there are no stable meson excitations in this case, a circumstance which is consistent with the fact that in the $E_7$ picture $g>0$ corresponds to the paramagnetic phase when confinement by magnetic field is absent. 

\section{Spectrum from TCSA}\label{sec:tcsa_comparison}

The theoretical predictions which we got from the semi-classical approximation can be checked by comparing them versus the numerical spectra obtained using the truncated conformal space approach, originally introduced by Yurov and Zamolodchikov \cite{1991IJMPA...6.4557Y}, and applied to the scaling region of the tricritical Ising model in \cite{LassigMussardoCardy1991}.  We omit a presentation of the TCSA and refer the reader to \cite{2018RPPh...81d6002J} for a general review of TCSA. Regarding our  implementation of the TCSA method, we have used the recently developed algorithm exploiting chiral factorisation of conformal field theory described in \cite{2022arXiv220106509H}, together with a renormalisation group improvement \cite{GiokasWatts,2015JHEP...09..146L} based on cut-off extrapolation. 

\subsection{Spectrum in the perturbed RSOS  $\mathcal{A}_3$ model}\label{subsec:A3TCSA}
\begin{figure}
	\centering
     \includegraphics[width=\columnwidth]{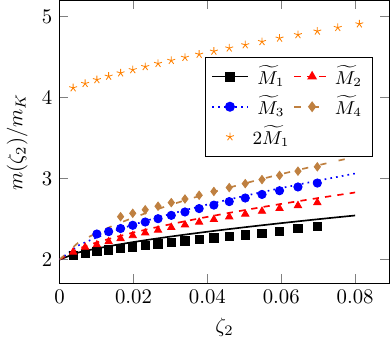}
	 \caption{TCSA results for the mesons in the thermally perturbed RSOS $\mathcal{A}_3$ model, with the semi-classical results shown as continuous lines. The orange curve shows the threshold line $2\widetilde{M}_1$: therefore the four mesons presented in this figure are all stable.}
	 \label{fig:mesonsA3}
\end{figure}
\begin{figure}[t]
	\centering
    \includegraphics[width=\columnwidth]{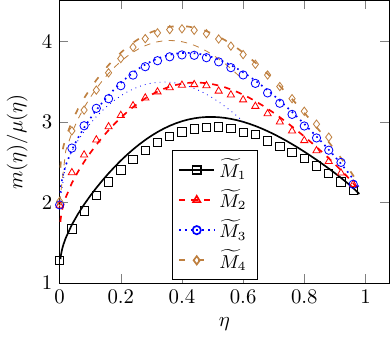}
	\caption{Particle and meson masses in the combined thermal and subleading magnetic perturbations. The masses $\widetilde{M}_i$ are labelled according to the meson masses following from the semi-classical predictions based on the $\mathcal{A}_3$ end-point $\eta=1$, shown by the thicker lines. Note that the first two mesons of the $\mathcal{A}_3$ confinement regime evolve into the $E_7$ breathers $A_2$ and $A_4$ as $\eta$ decreases, and that the RSOS $\mathcal{A}_3$ picture provides a good description of the mass spectrum for all the range as $\eta$ decreases from $1$ to $0$. The behaviour close to $\eta=1$ is presented in more details in Figure~\ref{fig:mesonsA3} with data corresponding to the interval $0.6<\eta<1$. The semi-classical predictions based on the $E_7$ end-point $\eta=0$ are shown with the thinner lines, and are only valid for sufficiently small $\eta$ (see Figure \ref{fig:mesonsigma} corresponding to the range $0<\eta<0.2$ for details). Note that the first two mesons of the $E_7$ picture evolve to the 3rd and 4th meson states of the RSOS  $\mathcal{A}_3$ description.}
	\label{fig:masscoupling_ferro_V}
\end{figure}
We start with the semi-classical regime around the RSOS $\mathcal{A}_3$ model (cf. Subsec. \ref{subsec:a3_confinement}). Using again the parameter 
\begin{equation}
    \zeta_2=\frac{\widetilde{\Delta\mathcal{E}}}{m_K^2}\,,
\end{equation}
the excellent agreement between the semi-classical predictions and the TCSA results is demonstrated in Fig. \ref{fig:mesonsA3}.

Note that for the model $\mathcal{A}\left[g,0,h'\right]$ describing the combined thermal and subleading magnetisation perturbation of the tricritical Ising model, there are two semi-classical descriptions depending on the relative size of the two couplings. Introducing the dimensionless parameter
\begin{equation}
\eta=\frac{{|h'|}^{9/5}}{{|h'|}^{9/5}+|g|^{9/8}}
\,\,\,,
\label{eq:etadef}
\end{equation}
the $E_7$ and RSOS $\mathcal{A}_3$ endpoints correspond to $\eta=0$ and $1$, respectively. It is also convenient to define our units in terms of the mass scale 
\begin{equation}
    \mu(\eta)=(1-\eta)m_1+\eta m_K\,,
\end{equation}
which interpolates smoothly between the $E_7$ mass gap $m_1$ \eqref{eq:e7gap} and the RSOS $\mathcal{A}_3$ kink mass $m_K$ \eqref{eq:a3gap}. The comparison of the two semi-classical pictures to the mass spectrum extracted from TCSA is shown in Fig. \ref{fig:masscoupling_ferro_V}.

Apart from the validity of the semi-classical approaches in their appropriate regimes $\eta\ll 1$ ($E_7$) and $(1-\eta)\ll 1$ ($\mathcal{A}_3$), it is remarkable that the semi-classical predictions of the $\mathcal{A}_3$ confinement continue to be quite accurate to very small values of $\eta$, down to $0.0025$. As we point out below, this means that the evolution of the mass spectrum can be described by semi-classical methods for the full range of parameters.

\subsection{Spectrum in the perturbed ferromagnetic $E_7$ model}

\begin{figure}[t]
	\centering
     \includegraphics[width=\columnwidth]{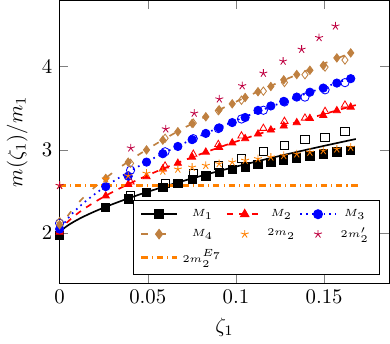}
	 \caption{TCSA results for the first four $11$ meson masses for the leading (filled markers) and the subleading (empty markers) magnetic perturbation of the $E_7$ model in the ferromagnetic phase, with the ($E_7$ based) semi-classical predictions shown with continuous lines. In this figure the mass  $m_2$ ($m'_2$) of the deformed $A_2$ particle in the $\sigma$ ($\sigma'$) perturbed theory is plotted in terms of the orange (purple) stars, while the original $E_7$ result ($2\times 2\cos(5\pi/18)$ by an orange line.}
	 \label{fig:mesonsigma}
\end{figure}

Now we turn to the semi-classical regime around the $E_7$ model, treated in Subsection \ref{subsec:e7_confinement}. The small parameter of the semi-classical approximation is 
\begin{equation}
    \zeta_1=\frac{\Delta\mathcal{E}}{m_1^2}\,,
\end{equation}
and the predicted masses only depend on $\zeta_1$ despite having two different perturbations. As demonstrated in Fig. \ref{fig:mesonsigma}, the TCSA data are fully consistent this prediction, with deviations only appearing for larger values of $\zeta_1$. In addition, the semi-classical approximation works better for higher meson states, as expected. While the data shown in Fig. \ref{fig:mesonsigma} only includes either pure $\sigma$ or $\sigma'$ perturbations, we have also verified that the agreement holds when both magnetic fields $h$ and $h'$ are nonzero. Note that $\eta=0.0025$ corresponds to $\zeta_1=0.012$, which is well inside the range of validity of the $E_7$ semi-classical regime as apparent from Fig. \ref{fig:mesonsigma}. As a result, the combination of the two semi-classical approaches provides a very good description of the evolution of the mass spectrum for the combined thermal and subleading magnetic deformation for the full range of parameters.

\begin{figure}[t]
    \centering
    \includegraphics[width=\columnwidth]{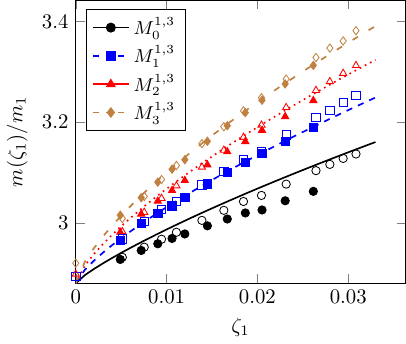}
    \caption{Meson level candidates obtained as deformation of $|A_1A_3\rangle$ scattering states in $E_7$. TCSA data are denoted by full markers for the $\sigma$, and empty markers for the $\sigma'$ perturbations. Note that the range of $\zeta_1$ is much smaller than in Fig. \ref{fig:mesonsigma}.}
    \label{fig:mesons13}
\end{figure}

The semi-classical approach predicts the existence of $\{1,3\}$ mesons formed as bound state of fundamental kink $A_1$ and the excited kink $A_3$. The mass of $A_3$ is $m_3=2m_1\cos(\pi/9)\approx 1.879 m_1$, and the $S$-matrix amplitudes entering \eqref{eq:nonidKquant} is
\begin{equation}
    S_{13}(\theta)=-f_{6}(\theta)f_{10}(\theta)f_{14}(\theta)\,.
\end{equation}
The results of the semi-classical approach are compared to TCSA in Fig. \ref{fig:mesons13}. Note that in this case the semi-classical predictions work in a much smaller range of $\zeta_1$ than for the bound states of two $A_1$ kinks, i.e., the $\{1,1\}$ mesons shown in Fig. \ref{fig:mesonsigma}.

The observation that the range of validity of the semi-classical approach is more limited for the $\{1,3\}$ mesons compared to the $\{1,1\}$ mesons can be understood as follows. As stressed in Sec.~\ref{sec:confinement}, the semi-classical approximation assumes that the meson states are well approximated by a combination of two-kink scattering states in the absence of the confining force i.e. $\Delta\mathcal{E}=0$. However, the confining perturbation breaks integrability and therefore generally opens inelastic channels mixing different kink configurations provided their hybridisation is kinematically allowed, which is always the case for two-kink states $A_1A_3$ mixing with $A_1A_1$ due to $m_3>m_1$. So, even though the fields $\sigma$ and $\sigma'$ are odd and then the mixing is a second order effect in $\zeta_1$, the inelastic channel that are kinematically allowed for the $\{1,3\}$ mesons implies a much smaller range of $\zeta_1$ for the semiclassical approximation.

The effect of inelastic channels can be detected numerically by computing the overlaps of the state vectors of the candidate meson levels at non-zero magnetic fields with the appropriate two-kink scattering states at the $E_7$ point with $h=h'=0$. Indeed, direct computations for $0.05<\zeta_1<0.15$ showed that the state vectors of $\{1,1\}$ meson level candidates are still dominated by $A_1A_1$ scattering states computed with zero magnetic field, both for $\sigma$ and $\sigma'$ perturbation, typically having a total overlap with the $A_1A_1$ scattering states in excess of $0.75$. Compared to the $\{1,1\}$ meson states, the $\{1,3\}$ meson level candidates turned out to have a much smaller total overlap with $A_1A_3$ scattering states, and for some $\{1,3\}$ meson states predicted by the semi-classical quantisation \eqref{eq:nonidKquant} it was not even possible to identify a candidate level in the TCSA spectrum. 

We remark that due to the density of energy levels of the spectrum, the identification of the $\{1,3\}$ meson states in the range depicted in Fig. \ref{fig:mesons13} had to be performed by a very careful process following the deformation of the spectral lines corresponding to $A_1A_3$ two-particle states as the coupling was gradually increased. This is in sharp contrast to the $\{1,1\}$ meson states, where the corresponding energy levels can be found by a simple look at the spectrum. In principle, the semi-classical approach predicts the existence $\{3,3\}$ meson states, as well as mesons involving the higher excited kink $A_6$. However, here we refrain from their examination for two main reasons. Firstly, the relevant part of  the TCSA spectrum is much higher in energy and therefore much denser, making identification of these states extremely difficult. Secondly, the inelastic processes invalidating the semi-classical approach are expected to be much stronger, and so the validity of the semi-classical predictions is expected to be restricted to even smaller values of $\zeta_1$.

\section{Conclusions}\label{sec:conclusions}

In this paper we have analysed the pattern of mesonic bound states which come from the confinement of the different kinks present in various perturbations of the tricritical Ising model. This model presents several novel aspects compared to previously studied cases such as perturbations of the critical Ising and Potts models. One of them is the presence of vacuum degeneracy which is not related to spontaneous symmetry breaking, leading to an asymmetric vacuum structure. As a result, the confinement dynamics resulting from perturbing the $\mathcal{A}_3$ endpoint depends on the sign of the perturbing (thermal) coupling. In addition, in contrast to confinement resulting from explicit symmetry breaking fields, the resulting perturbation smoothly interpolates to another endpoint ($E_7$) with (symmetry breaking) degenerate vacua. Another interesting feature of the confinement dynamics induced by the thermal perturbation of the $\mathcal{A}_3$ end point is that the validity of semi-classical approximation extends well beyond weak coupling, and eventually describes the spectrum very well even close to the other ($E_7$) endpoint, for which we have no compelling theoretical explanation at present. Furthermore, the $E_7$ endpoint admits excited kinks, for which we derived the appropriate semi-classical quantisation conditions, which were verified to describe the spectrum; however, these predictions had a more restricted range of validity, which can be understood by the larger contribution of inelastic channels to the kink-kink scattering. Since the interaction which induces the confinement of the kinks breaks the integrability of the original model, the mesonic excitations with mass above the appropriate kinematic threshold are expected to have a finite lifetime. However, for the case of confinement in the Ising model it was argued that even excitations above the threshold could have anomalously long lifetimes \cite{2005PhRvL..95y0601R,2019PhRvL.122m0603J,2019PhRvB..99s5108R}. In view of these results, the issue of the decay widths of the mesonic excitations in the tricritical Ising  model is an interesting question left open for future investigations.

\subsection*{Acknowledgments}
We are grateful to A. Milsted for pointing out some related results in Ref.~\cite{2020arXiv201207243M}. GM acknowledges the grant Prin $2017$-FISI. The work of ML was supported by the National Research Development and Innovation Office of Hungary under the postdoctoral grant PD-19 No. 132118. GT and ML were partially supported by the  National Research, Development and Innovation Office (NKFIH) through the OTKA Grant K 138606 and also within the Quantum Information National Laboratory of Hungary. This work was also partially supported by the CNR/MTA Italy-Hungary 2019-2021 Joint Project ``Strongly interacting systems in confined geometries”. 

\providecommand{\href}[2]{#2}\begingroup\raggedright\endgroup

\bibliographystyle{utphys}

\end{document}